\documentclass[onecolumn,floatfix,superscriptaddress,showpacs,showkeys,nofootinbib]{revtex4}

\usepackage{epsfig}
\usepackage{amssymb,latexsym,amsmath}

\begin{document}

\vspace{1cm}

\title{CRITICAL LINE OF THE 
DECONFINEMENT PHASE TRANSITIONS 
}

\vspace{0.5cm}

\author{ Mark I. Gorenstein
\footnote{
The results of this paper were presented as 
invited talk at the 2nd International Workshop on the Critical
Point and Onset of the Deconfinement, 30 March - 3 April 2005, Bergen, Norway} 
}
\affiliation{ Bogolyubov Institute for Theoretical Physics, Kiev, Ukraine}
 \affiliation{Frankfurt Institute for Advanced Studies, Frankfurt, Germany}
 \affiliation{Institut f\"ur Theoretische Physik, Johann Wolfgang Goethe Universit\"at,
 Frankfurt, Germany}
 \author{Marek Ga\'zdzicki
\footnote{On leave of absence from \'Swi\c{e}tokrzyska Academy, Kielce, Poland}
}
 \affiliation{Institut f\"ur Kernphysik, Johann Wolfgang Goethe Universit\"at Frankfurt, Germany}
\author{Walter Greiner}
\affiliation{ Frankfurt Institute for Advanced Studies, Frankfurt, Germany}
 \affiliation{Institut f\"ur Theoretische Physik, Johann Wolfgang Goethe Universit\"at,
 Frankfurt, Germany}

\vspace{0.5cm}

\begin{abstract}
Phase diagram of strongly interacting matter is discussed 
within the exactly solvable 
statistical model of the quark-gluon bags. The model predicts two phases
of matter: the hadron gas at a low temperature $T$ and baryonic chemical potential
$\mu_B$, and the quark-gluon gas at a high $T$ and/or $\mu_B$. 
The nature of the phase transition
depends on a form of the bag mass-volume spectrum (its pre-exponential factor),
which is expected to change with the $\mu_B/T$ ratio.
It is therefore likely that
the line of the 1$^{st}$ order transition at a high $\mu_B/T$ ratio
is followed by the line
of  the 2$^{nd}$ order phase transition at an intermediate $\mu_B/T$, and then by the lines
of "higher order transitions" at a low $\mu_B/T$. 
\end{abstract}


\maketitle

\section{Introduction}

Investigation of the properties of strongly interacting matter
at a high energy density is one of the most important subjects
of the contemporary physics.
In particular, the hypothesis that at high energy densities
the matter is in the form of quark-gluon plasma (QGP) \cite{qgp} rather
than a gas of hadrons (HG) motivated a first stage of the broad 
experimental program of study of ultra-relativistic nucleus--nucleus
collisions \cite{exp}. 
Over the last 20 years  rich data were collected by experiments
located at Alternating Gradient Synchrotron (AGS) and 
Relativistic Heavy Ion Collider
(RHIC) in Brookhaven National Laboratory, USA and
at Super Proton Synchrotron (SPS)  in CERN, Switzerland.
The results indicate that the properties of the created matter
change rapidly in the region of the low SPS energies \cite{onset} 
($\sqrt{s}_{NN} \approx$ 6--12 GeV).
The observed effects confirm predictions for the transition from 
hadron gas to QGP \cite{gago} and thus indicate that in fact 
a new form of strongly interacting matter exists in nature at
sufficiently high energy densities.

What are the properties of the transition between the two phases
of strongly interacting matter?
This question motivates the second stage of the investigation
of nucleus-nucleus collisions.
A new experimental heavy ion program  
($\sqrt{s}_{NN} \approx$ 5--20 GeV) at the CERN SPS is in preparation
\cite{loi} 
and a possible study at the BNL RHIC is under discussion.

Based on the numerous examples of the well-known substances
a conjecture was formulated (see review paper \cite{misha}
and references there in)
that the transition from HG to QGP is a 1$^{st}$ order
phase transition at low values of a temperature $T$ and a high baryo-chemical
potential $\mu_B$ and it is a rapid cross--over at a high $T$ and
a low $\mu_B$.
The end point of the 1$^{st}$ order phase transition line is expected
to be the 2$^{nd}$ order critical point. 
This hypothesis seems to be supported by  
QCD--based qualitative considerations \cite{misha} and
first semi--quantitative lattice QCD calculations \cite{fodorkatz}. 
The properties of the deconfinement phase transition are, however,
far from being well established.
This stimulates our study of the 
transition domain  within the statistical model of quark-gluon bags.

The bag model \cite{bag} was invented in order to describe the hadron spectrum, 
the hadron masses and their proper volumes. 
This model is also successfully used for
a description of the deconfined phase (see e.g. Ref.~\cite{bag-qgp}).
Thus the model suggests a possibility for
a unified treatment of both phases. Any configuration
of the system and, therefore, each term in the
system partition function, can be regarded as a many-bag state
both below and above the transition domain. 
The statistical model 
of quark-gluon bags discussed in this paper  contains in itself two
well-known models of deconfined and confined phases:
the bag model of the QGP \cite{bag-qgp} and
the hadron gas models \cite{hg}. They are surprisingly
successful in describing the bulk properties of hadron
production in high energy nuclear collisions and 
thus one may  hope that  the 
model presented here may reflect the basic features of nature in the
transition domain.

The paper is organized as follows.
The important properties of the statistical bootstrap model with
the van der Waals repulsion are summarized in Sect.~II.
In Sect.~III the statistical model of the quark--gluon bags is
presented.
Within this model properties of the transition region 
for $\mu_B=0$
are studied in
Sect.~IV and the analysis is extended to the complete $T-\mu_B$ plane
in Sect.~V.
The paper is closed by the summary and outlook given in Sect. VI.

\section{Statistical Bootstrap Model and Van der Waals Repulsion}
The grand canonical partition function for an ideal Boltzmann gas of particles of
mass $m$ and a number of internal degrees of freedom (a degeneracy factor) 
$g$, in a volume $V$ and at a temperature $T$ is given by: 
\begin{align}\label{Zid}
Z(V,T)~ = ~
\sum_{N=0}^{\infty}~\frac{1}{N!}~\prod_{j=1}^{N}\int\frac{gV\;d^3k_j}{(2\pi)^3}
 \exp\left[~-~\frac{(k_j^2~+~m^2)^{1/2}}{T}
~\right]~
=~\sum_{N=0}^{\infty}\frac{[V~g\phi(T,m)]^N}{N!}~=~\exp[Vg\phi(T,m)]~,
\end{align}
where
\begin{align}\label{phim}
\phi(T,m)~  \equiv ~\frac{1}{2\pi^2}~\int_0^{\infty}k^2dk~
\exp\left[-~\frac{(k^2~+~m^2)^{1/2}}{T} \right]
~=~\frac{m^2T}{2\pi^2}~K_2\left(\frac{m}{T}\right)
\end{align}
and $K_2$ is the modified Bessel function.
The function $\phi(T,m)$ is equal to the particle number density:
\begin{align}
n(T)~\equiv~\frac{\overline{N}(V,T)}{V}~=~g\phi(T,m)~.
\end{align}
The ideal gas pressure and energy density can be derived from Eq.~(\ref{Zid}) as: 
\begin{align}\label{pepsilon}
p(T)~ \equiv ~T~\frac{\ln Z(V,T)}{V}~=~ T~g\phi(T,m) ~,~~~~
 \varepsilon(T)~ \equiv ~T~\frac{dp}{dT} ~-~ p(T)~=~
T^2~g\frac{d\phi(T,m)}{dT}~. 
\end{align}

One can easily generalize the ideal gas formulation (\ref{Zid}) to the mixture of particles with
masses $m_1,...,m_n$  and degeneracy factors $g_1,...,g_n$:
\begin{align}\label{Zsbm}
 Z(V,T)~ = ~ \sum_{N_1=0}^{\infty}~...~\sum_{N_n=0}^{\infty}
\frac{\left[Vg_1\phi(T,m_1)\right]^{N_1}}{N_1!}~...
~ \frac{\left[Vg_n\phi(T,m_n)\right]^{N_n}}{N_n!}
~=~\exp\left[V~\sum_{j=1}^{n}g_j\phi(T,m_j)\right]~. 
\end{align}
The sum over different particle species $j$ can be extended to infinity.
It is convenient to introduce the mass spectrum density $\rho(m)$, so that 
$\rho(m)dm$ gives the number of different particle mass states in the interval
$[m,m+dm]$, i.e.
$\sum_{j=1}^{\infty} g_j...=
\int_0^{\infty}dm...\rho(m) 
$.
In this case the pressure and the energy density are given by: 
\begin{align}
p(T)~=~ T~\int_0^{\infty}dm~\rho(m)~\phi(T,m)~,
~~~~
\varepsilon(T)~ =~
T^2~\int_0^{\infty}dm~\rho(m)~\frac{d\phi(T,m)}{dT}~. \label{esbm}
\end{align}

\noindent
Eqs. (\ref{esbm}) were introduced by Hagedorn \cite{Hag} for the 
mass spectrum increasing exponentially for $m\rightarrow\infty$:
\begin{align}\label{rhom}
\rho(m)\simeq ~C~m^{-a}~\exp(bm)~, 
\end{align}
where $a$, $b$ and $C$ are model parameters. 
This form of the  spectrum (\ref{rhom}) was further derived from  
the statistical bootstrap model \cite{Frau}. 
It can be shown, that within this model the temperature $T_H\equiv 1/b$ 
(the ''Hagedorn temperature'') is the maximum temperature of the matter.
%
The behavior of thermodynamical functions (\ref{esbm}) with the mass spectrum
(\ref{rhom}) depends crucially on the parameter $a$.
In particular,  
in the limit  $T\rightarrow T_H$ the pressure and the energy density approach:
\begin{align}
 & p~,~\varepsilon~\rightarrow~\infty~,~~~~~~~~~~~~~~~~~~~~~{\rm for}~~~a~\le~\frac{5}{2}~;
\label{a1} \\
 & p~\rightarrow const~,~ \varepsilon~\rightarrow~\infty~,~~~~~~~~{\rm for}~~~
 \frac{5}{2}~\le~a~\le~\frac{7}{2}~; \label{a2}
 \\
 & p~,~\varepsilon~\rightarrow~const~,~~~~~~~~~~~~~~~~~{\rm for}~~~a~>~\frac{7}{2} ~.\label{a3}
\end{align}
Up to here
all particles including those with $m\rightarrow\infty$ were treated as point-like objects.
Clearly this is an unrealistic feature of the statistical bootstrap
model. 
It can be overcome by introduction of hadron proper volumes which 
simultaneously mimic
the repulsive interactions between hadrons. 
The van der Waals excluded volume procedure \cite{pt} can be applied for this purpose
\cite{pt} (other approaches were discussed in Refs.~\cite{HR,K}).
The volume $V$ of the ideal gas (\ref{Zid})
is substituted  by the "available volume" $V-v_oN$, where
$v_o$ is a parameter which corresponds to a particle proper volume. 
The partition function then reads:
\begin{align}\label{Zvdw}
Z(V,T)~=~\sum_{N=0}^{\infty}\frac{[(V~-~v_oN)~g \phi(T,m)]^N}{N!}~
\theta(V-v_oN)~.
\end{align}
The pressure of the van der Waals gas can be calculated from the
partition function  (\ref{Zvdw}) by use of its Laplace transform.
This procedure is necessary because the "available volume", $V-v_oN$,
depends on the particle number $N$.
The Laplace transform of  (\ref{Zvdw}) is \cite{pt,vdw1}:
\begin{align}\label{Zs}
& \hat{Z}(s,T)~\equiv\int_0^{\infty}dV\exp(-sV)~Z(V,T)
~=~\sum_{N=0}^{\infty}\frac{[g \phi(T,m)]^N}{N!}
\int_{v_oN}^{\infty}dV~(V~-~v_oN)^N~\exp(-sV)
\nonumber\\
&=~\frac{1}{s}~\sum_{N=0}^{\infty}\left[\frac{\exp(-v_os)~g\phi(T,m)}{s}\right]^N~
=~\left[~s~-~\exp(-v_os)~g \phi(T,m)\right]^{-1}~.
\end{align}
In the thermodynamic limit, $V\rightarrow\infty$, the partition function 
behaves as
$Z(V,T)\simeq \exp\left[pV/T \right]$.
An exponentially increasing $Z(V,T)$ generates the farthest-right singularity $s^*=p/T$ 
of the function $\hat{Z}(s,T)$ in variable $s$. 
This is because the integral over $V$ in Eq.~(\ref{Zs}) 
diverges at its upper limit for $s < p/T$. 
Consequently, the pressure can be expressed as
\begin{align}\label{p-s}
p(T)~=~T~\lim_{V\rightarrow\infty}\frac{\ln Z(V,T)}{V}~=~T~s^*(T)~,
\end{align}
and the farthest-right singularity 
$s^*$ of $\hat{Z}(s,T)$ (\ref{Zs}) can be calculated from the transcendental
equation \cite{pt,vdw1}:
\begin{align}\label{s*vdw}
s^*(T)~=~\exp\left[- ~v_o s^*(T)\right]~g \phi(T,m)~.
 \end{align}
Note that the singularity $s^*$ is not related to 
phase transitions in the system. Such a singularity exists for any 
statistical system.   
For example, for the ideal gas ($v_o=0$ in  Eq.~(\ref{s*vdw}))
$s^*=g\phi(T,m)$ and thus from  Eq.~(\ref{p-s}) one gets
$p=Tg\phi(T,m)$ which corresponds to the ideal gas equation of state
(\ref{pepsilon}). 

\section{Gas of Quark-Gluon Bags}
The van der Waals gas consisting of $n$ hadronic  species,
which are called bags in what follows. is considered in this section.
Its partition function reads:
\begin{align}
 Z(V,T)~ &=  ~\sum_{N_1=0}^{\infty}~...~\sum_{N_n=0}^{\infty}
\frac{\left[\left(V-v_1N_1-...-v_nN_n\right)~g_1\phi(T,m_1)\right]^{N_1}}{N_1!}
~\times ... 
 \nonumber \\
& ...~\times~
\frac{\left[\left(V-v_1N_1-...-v_nN_n\right)~g_n\phi(T,m_n)\right]^{N_n}}{N_n!}
~ \theta\left(V-v_1N_1-...-v_nN_n\right)~, 
\label{qqq}
\end{align}
where $(m_1,v_1), ..., (m_n,v_n)$ are the masses and volumes of the bags.
The Laplace transformation of Eq.~({\ref{qqq}) gives
\begin{align}\label{Zsn}
 \hat{Z}(s,T)~=~\left[~s~-~\sum_{j=1}^n \exp\left(-v_j s\right)~g_j\phi(T,m_j)\right]^{-1}~.
 \end{align}
As long as the number of bags, $n$, is finite, the only possible singularity 
of $\hat{Z}(s,T)$ (\ref{Zsn}) is its  pole. 
However, in the case of an infinite number of bags the second singularity of
$\hat{Z}(s,T)$ may appear. This case is discussed in what follows.

Introducing the bag mass-volume spectrum, $\rho(m,v)$, so that
$\rho(m,v)dmdv$ gives the number of bag states in the mass-volume 
interval $[m,v;m+dm,v+dv]$, the sum over different bag states in
definition of $Z(V,T)$ can be  replaced  by the integral,
$\sum_{j=1}^{\infty}g_j ...=\int_0^{\infty}dm dv ...\rho(m,v)$.  
Then, the Laplace transform of $Z(V,T)$ reads \cite{pt}:
\begin{align}\label{Zsbag}
 \hat{Z}(s,T)~&\equiv\int_0^{\infty}dV\exp(-sV)~Z(V,T)
~=~\left[~s~-~f(T,s)\right]^{-1}~,
\end{align}
\noindent
where
\begin{align}
  f(T,s)~&=~  \int_0^{\infty} dmdv
~\rho(m,v)~\exp(-vs)~\phi(T,m) ~.\label{fs}
\end{align}
The  pressure is again given by the farthest-right singularity: $p(T)=Ts^*(T)$.
One evident singular point of $\hat{Z}(s,T)$ (\ref{Zsbag}) is  the pole singularity,
$s_H(T)$: 
\begin{align}\label{sH}
s_H(T)~=~f\left(T,s_H(T)\right)~.
\end{align}
As mentioned above this is the only singularity of $\hat{Z}(s,T)$ if one restricts
the mass-volume bag spectrum to a finite number of states. 
For an infinite number of mass-volume states 
the second singular point of  $\hat{Z}(s,T)$ (\ref{Zsbag}), $s_Q(T)$,
can emerge, which is due to a possible singularity of the function $f(T,s)$ (\ref{fs}) itself.
The system pressure takes then the form:
\begin{align}\label{ps*}
p(T)~=~Ts^*(T)~=~T\cdot max\{s_H(T),s_Q(T)\} ~,
\end{align}
and thus 
the farthest-right singularity $s^*(T)$ of $\hat{Z}(s,T)$ (\ref{Zsbag})
can be either the pole singularity $s_H(T)$ (\ref{sH}) or 
the $s_Q(T)$ singularity of the function $f(T,s)$ (\ref{fs}) itself.
The mathematical mechanism for possible phase transition (PT)
in our model is the "collision" of the two singularities,
i.e. $s_H(T)=s_Q(T)$ at PT temperature $T=T_C$ 
(see Fig.~\ref{fs1})~\footnote{Note that 
the same technique has been recently used to describe the liquid-gas PT connected to the
multi-fragmentation phenomena in 
intermediate energy nucleus-nucleus collisions \cite{bug1,bug2}.}. 
In physical terms this can be interpreted as the existence of
two phases of matter, namely, the hadron gas with the pressure,  $p_H=Ts_H(T)$,
and the quark gluon plasma with the pressure $p_Q=Ts_Q(T)$. At a given temperature $T$
the system prefers to stay in a phase
with  the higher pressure. 
The pressures of both phases are equal at the PT temperature $T_C$. 

An important feature of this  modeling of the
phase transition should be stressed here.
The transition, and thus the occurrence of the two phases
of matter, appears as a direct consequence of the postulated
general partition function (a single equation of state).
Further on, the properties of the transition, e.g. its location
and order, follow from the partition function and  are
not assumed. 
This can be confronted  with the well-known phenomenological
construction of the phase transition, in which the existence of
the two different phases of matter and the nature of the transition between
them are postulated.

The crucial ingredient of the  model presented here which defines
the presence, location and the order of the PT
is the form of the mass-volume spectrum of bags $\rho(m,v)$. 
In the region where both  
$m$ and $v$ are large it can be described within the bag model \cite{bag}.
In the simplest case of a bag filled with the non-interacting massless quarks and gluons
one finds \cite{pt1}:
\begin{align}\label{rhomv}
 \rho(m,v)~\simeq
 ~C~v^{\gamma}(m-Bv)^{\delta} ~\exp\left[\frac{4}{3}~\sigma_Q^{1/4}~
 v^{1/4}~(m-Bv)^{3/4}\right]~,
\end{align}
where $C$, $\gamma$, $\delta$ and $B$ 
(the so-called bag constant, $B\approx 400$~MeV/fm$^3$~\cite{bag-qgp}) are the model parameters and
\begin{align}\label{sigmaQ}
 \sigma_Q~=~\frac{\pi^2}{30}\left(g_g~+~\frac{7}{8}g_{q\bar{q}}\right)~
 =~
\frac{\pi^2}{30}\left(2\cdot 8~+~\frac{7}{8}\cdot 2\cdot 2\cdot 3
\cdot 3\right)~=~\frac{\pi^2}{30}~\frac{95}{2}
\end{align}
is the Stefan-Boltzmann constant counting gluons (spin, color) and (anti-)quarks
(spin, color and $u$, $d$, $s$-flavor) degrees of freedom.
This is the asymptotic expression assumed to be valid for a sufficiently large 
volume and mass of a  bag, $v>V_0$ and $m>Bv+M_0$. The validity limits
can be estimated to be $V_0\approx 1$~fm$^3$ and $M_0\approx 2$~GeV \cite{pt2}.
The mass-volume spectrum function:
 \begin{align}
 \rho_H(m,v)~=~\sum_{j=1}^n~ g_j~ \delta(m-m_j)~\delta(v-v_j)
\end{align}
should be added to $\rho(m,v)$ in order to reproduce the known low-lying hadron states 
located at $v < V_0$ and $m < BV_0+M_0$.
The mass spectra of the resonances are described by the
Breit-Wigner functions. Consequently, a
general form of  $f(T,s)$ (\ref{fs}) reads:
\begin{align}\label{fsHQ}
 f(T,s)~\equiv~f_H(T,s)+f_Q(T,s)
 =\sum_{j=1}^n g_j\exp(-v_js)~\phi(T,m_j)
 +\int_{V_o}^{\infty}dv\int_{M_o+Bv}^{\infty}dm~\rho(m,v)\exp(-sv)\phi(T,m)~,
\end{align}
where $\rho(m,v)$ is given by Eq.~(\ref{rhomv}).

The behavior of $f_Q(T,s)$ is  discussed in the following.
The integral over mass  in Eq.~(\ref{fsHQ}) can be calculated by the steepest 
decent estimate. Using the asymptotic expansion of the $K_2$-function
one finds $\phi(T,m)\simeq (mT/2\pi)^{3/2}\exp(-m/T)$
for $m\gg T$.
 A factor exponential in $m$ in the last term of Eq.~(\ref{fsHQ}) is given by: 
\begin{align}
  \exp\left[-\frac{m}{T}~+~\frac{4}{3}\sigma_Q^{1/4}v^{1/4}
 (m-Bv)^{3/4}\right]~\equiv~\exp[U(m)]~.
\end{align}
The function $U(m)$ has a maximum at:
\begin{align}
  m_0~=~\sigma_QT^4v~+~Bv~.
\end{align}
Presenting $U(m)$ as
\begin{align}
U(m)~\simeq~U(m_0)~ +~ \frac{1}{2}~\left(\frac{d^2U}{dm^2}\right)_{m=m_0}~(m-m_0)^2~
=~vs_Q(T)~-~ \frac{(m-m_0)^2}{8\sigma_Q~vT^5}~,
\end{align}
with
\begin{align}\label{sQ}
s_Q(T)~\equiv~ \frac{1}{3}~\sigma_Q~T^3~-~\frac{B}{T}~,
\end{align}
one finds
\begin{align}\label{fQs}
 f_Q(T,s)~\simeq~u(T)
 ~  \int_{V_0}^{\infty}dv~v^{2+\gamma +\delta}~\exp\left[-v\left(s~-s_Q(T)\right)\right]~,
 \end{align}
 where
 $u(T)=C\pi^{-1}\sigma_Q^{\delta+1/2}~T^{4+4\delta}~\left(\sigma_QT^4+B\right)^{3/2}$.
The function $f_Q(T,s)$ (\ref{fQs}) has the singular point $s=s_Q$ 
because for $s<s_Q$ the integral over $dv$  diverges at its upper limit. 

The first term of Eq.~(\ref{fsHQ}), $f_H$, represents the contribution of a finite number of low-lying
hadron states. This function has no $s$-singularities at
any temperature $T$. The integration over the region $m,v \rightarrow\infty$
generates the singularity $s_Q(T)$ (\ref{sQ}) of the $f_Q(T,s)$ function (\ref{fQs}). 
As follows from the discussion below
this singularity should be  associated with the QGP phase. 

By construction the function $f(T,s)$ (\ref{fs}) is a positive one, so
that $s_H(T)$ (\ref{sH}) is also positive. 
On the other hand, it can be seen from Eq.~(\ref{sQ}) that 
at a low $T$ $s_Q(T)<0$. 
Therefore,
 $s_H > s_Q$ at small $T$, and according to Eq.~(\ref{ps*}) it follows:
\begin{align}\label{psH}
p(T)~=~T~ s_H~,~~~~ \varepsilon(T)~=~T^2~\frac{ds_H}{dT}~.
\end{align}
The system of the quark-gluon bags is in a hadron phase.

If  two singularities "collide", $T=T_C$ and $s_H(T_C)=s_Q(T_C)$, the singularity $s_Q(T)$
can become the farthest-right singularity at $T>T_C$.
In Fig.~\ref{fs1} the  dependence of the function $f(T,s)$ on
$s$ and its singularities are sketched for $T_1 < T_2=T_C < T_3$. 
The thermodynamical functions 
defined by the singularity $s_Q(T)$ are:
\begin{align}\label{psQ}
p(T)~=~T~ s_Q~=~\frac{\sigma_Q}{3}~T^4~-~B~,~~~~
\varepsilon(T)~=~T^2~\frac{ds_Q}{dT}~=~\sigma_Q~T^4~+~B~ 
\end{align}
and thus they describe the QGP phase. 

\begin{figure}[h!]
 \hspace{-2.1cm}
\epsfig{file=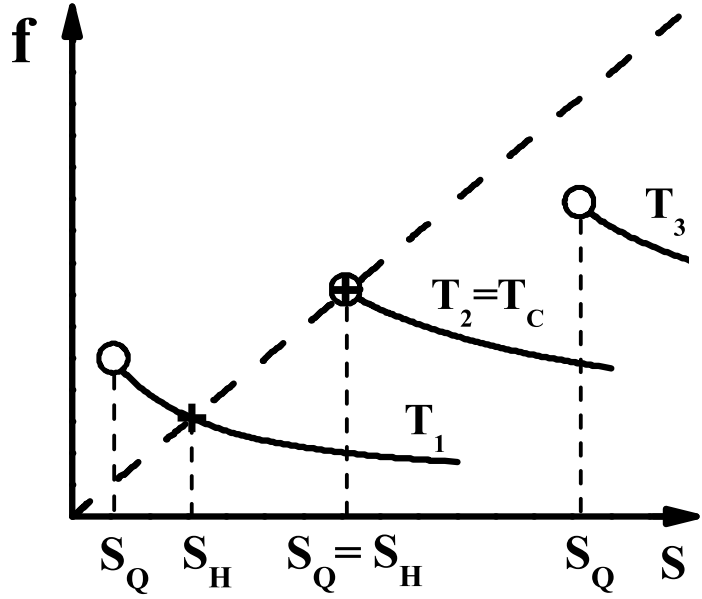,height=8cm,width=10cm}
\caption{The dependence of  $f(T,s)$ on $s$ for three different temperatures:
$T_1<T_2=T_C<T_3$ (solid lines). The pole singularity $s_H$ and the singularity $s_Q$ are denoted
by circles and crosses, respectively. A PT corresponds to the "collision" of
two singularities $s_H=s_Q$ at the temperature $T_C$.
\label{fs1}}
\end{figure}

The existence and the order of the phase transition depend
on the values of the parameters of the model.
From Eq.~(\ref{fQs}) follows that a PT exists, i.e. $s^*=s_Q$ at high $T$, 
provided
$\gamma+\delta < -3$ 
(otherwise $f(T,s_Q)=\infty$, and
$s_H>s_Q$ for all $T$, for illustration see Fig.~\ref{fs3}, left).

\begin{figure}[h!]
\hspace{-0.1cm}
\epsfig{file=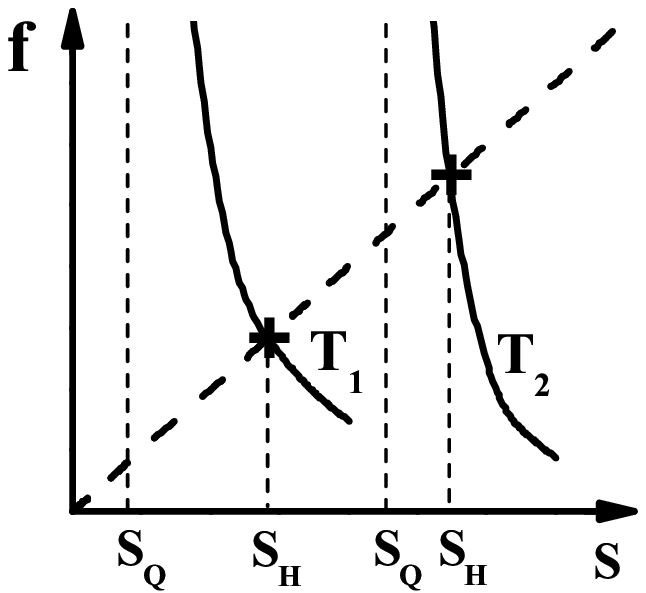,height=6.0cm,width=7cm}\hspace{1.5cm}
\epsfig{file=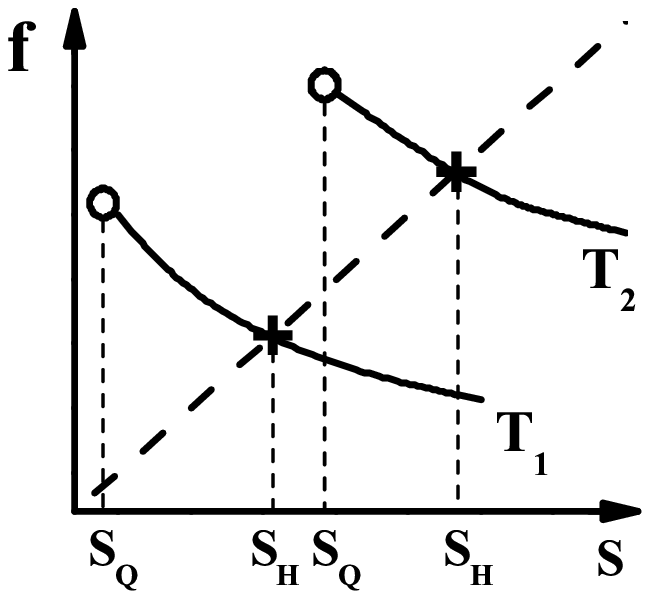,height=6.0cm,width=7cm}
\caption{
The dependence of  $f(T,s)$ on $s$ for two different temperatures,
$T_1<T_2$ (solid lines).
The figures demonstrate the absence of 
the PT in the model if $f(T,s_Q)=\infty$
(left), or if $f(T,s_Q)>s_Q$ at all $T$ (right).  
\label{fs3}}
\end{figure}
%
%
%
In addition it is required that 
$f(T,s_Q)<s_Q(T)$ at $T\rightarrow\infty$,
(otherwise $s^*=s_H>s_Q$ for all $T$, for illustration see Fig.~\ref{fs3}, right).
For $\gamma+\delta < -3$ one finds $f_Q(T,s_Q)\propto
T^{10+4\delta}$ at 
%
%
$T\rightarrow\infty$. 
On the other hand, 
$ s_Q(T) \propto T^3$ at a high $T$ and therefore
$\delta <-7/4$.
Consequently the general conditions for the existence of any
phase transition in the model are: 
\begin{align}\label{gamdel}
\gamma~<~-~\frac{5}{4}~,~~~~~~~\delta
~<~-~\frac{7}{4}~.
\end{align}
The parameters $\gamma$ and $\delta$ can be calculated within the model
provided constraints defining allowed quark-gluon bags are given
\cite{pt1, pt2}.
Assuming massless non-interacting quarks and gluons with the condition of 
the total three-momentum vector equal to zero one gets \cite{pt1}:
\begin{align}\label{gamdel1}
\gamma~=~-~\frac{5}{4}~,~~~~~~~\delta
~=~-~\frac{7}{4}~,
\end{align}
and consequently there is no phase transition in the model.
For a graphical illustration of this case see Fig.~\ref{fs3} left.
An additional selection of only colorless bags yields \cite{pt2}:
\begin{align}\label{gamdel2}
\gamma~=~-~\frac{9}{4}~,~~~~~~~\delta
~=~-~\frac{19}{4}~,
\end{align}
so that the conditions (\ref{gamdel}) are fulfilled and the model
leads to the phase transition, see Fig.~1 for illustration.
Based on these two simple examples one concludes that the
parameters $\gamma$ and $\delta$ depend significantly  on the
set of selected constraints. These constraints in statistical mechanics
are usually ignored as  the related to them pre-exponential factors
do not contribute to the equation of state in the thermodynamic
limit. Because of this possible sets of physical constraints 
are not well defined 
and their consequences were not studied.
Therefore
in what follows $\gamma$ and $\delta$ are
considered as free model parameters.

\section{1$^{st}$, 2$^{nd}$ and higher ORDER PHASE TRANSITIONS}
%
In this section the order of the PT in the system of
quark-gluon bags is discussed.

The 1$^{st}$ order PT takes place at $T = T_C$
provided:
\begin{align}\label{pt11}
 s_Q(T_C)~=~s_H(T_C)~,~~~~
  \left(\frac{ds_Q}{dT}\right)_{T=T_C}~
  > ~ \left(\frac{ds_H}{dT}\right)_{T=T_C}~.
 \end{align}
Thus the energy density ($\varepsilon = T^2 ds/dT$) has discontinuity
(latent heat) at the 1$^{st}$ order PT.
Its dependence on $T$ is shown in 
Fig.~\ref{pt2}, left. 
In calculating $ds_H/dT$ it is important to note that  the function $s_H(T)$ (\ref{sH}) is 
only defined  for $T\le T_C$, i.e. for $s_H(T)\ge s_Q(T)$.
%

\begin{figure}[h!]
 \hspace{-0.1cm}
\epsfig{file=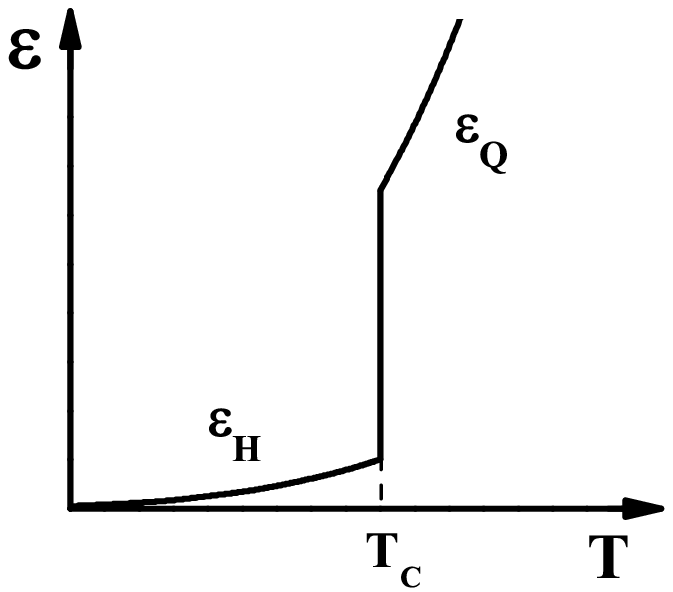,height=8.cm,width=7cm}\hspace{1.5cm}
\epsfig{file=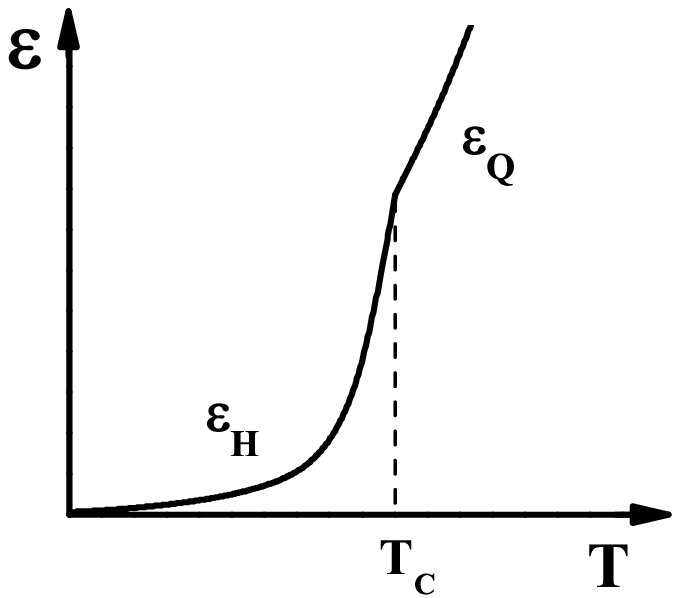,height=8.cm,width=7cm}
\caption{
The dependence of the energy density $\varepsilon$ on temperature for the 
1$^{st}$ (left) and 2$^{nd}$ (right) order PT.
The energy density $\varepsilon(T)$ has a discontinuity at $T=T_C$ 
for the 1$^{st}$ order PT 
whereas for the 2$^{nd}$ order PT $\varepsilon(T)$ is a continuous
function of $T$, but $d\varepsilon/dT$ has a discontinuity at $T=T_C$.  
\label{pt2}}
\end{figure}
%
%

The 2$^{nd}$ order PT takes place at $T = T_C$ provided:
\begin{align}\label{pt22}
 s_Q(T_C)~=~s_H(T_C)~, ~~~~ 
  \left(\frac{ds_Q}{dT}\right)_{T=T_C}~
  = ~ \left(\frac{ds_H}{dT}\right)_{T=T_C}~,~~~~
 \left(\frac{d^2s_H}{d^2T}\right)_{T=T_C}~
  \ne ~ \left(\frac{d^2s_Q}{d^2T}\right)_{T=T_C}~.
\end{align}
Hence the energy density is a continuous function of $T$, but its
first derivate has a discontinuity, for illustration see Fig.~\ref{pt2}, right.


What is the order of the PT in the model?
To answer this question  Eq.~(\ref{sH}) should be rewritten as:
\begin{align}\label{sH1}
 s_H~=~f_H~+~u~\int_{V_o}^{\infty}dvv^{-\alpha}\exp\left[-v\left(s_H-s_Q\right)
 \right]~,
 \end{align}
 where
 $\alpha\equiv-(\gamma+\delta+2)~>1$.
Differentiating both sides with respect to $T$ (the prime denotes the $d/dT$ derivative)
one gets:
\begin{align}\label{sHprime}
 s_H^{\prime}~=~f_H^{\prime}~+~u^{\prime}~
 \int_{V_o}^{\infty}dvv^{-\alpha}~
 \exp\left[-v\left(s_H-s_Q\right)
 \right]~+u~\int_{V_o}^{\infty}dvv^{-\alpha+1}~\left(s_Q^{\prime}-s_H^{\prime}\right)~
 \exp\left[-v\left(s_H-s_Q\right)
 \right]~,
\end{align}
from which follows:
\begin{align}\label{sHprime1}
s_H^{\prime}~=~\frac{G~+~F\cdot  s_Q^{\prime}}{1~+~F}~,
\end{align}
where
\begin{align}\label{GF}
 G~\equiv~
f_H^{\prime}+u^{\prime}\int_{V_o}^{\infty}dvv^{-\alpha}
\exp\left[-v\left(s_H-s_Q\right)
 \right] ~,~~~~
 F~\equiv ~u~
\int_{V_o}^{\infty}dvv^{-\alpha+1}\exp\left[-v\left(s_H-s_Q\right)
 \right]~.
\end{align}

It is easy to see that the transition is of the 1$^{st}$ order,
i.e. $s_Q^{\prime}(T_C)>s_H^{\prime}(T_C)$, provided $\alpha>2$.
The 2$^{nd}$ or higher order phase transition takes place
provided $s_Q^{\prime}(T_C)=s_H^{\prime}(T_C)$ at $T=T_C$.
This condition is satisfied  when $F(T)$ diverges to infinity
at $T\rightarrow (T_C-0)$, i.e. for $T$ approaching $T_C$ from below.

One notes that
the exponential factor in $\rho(m,v)$ (\ref{rhomv}) generates the singularity $s_Q(T)$ (\ref{sQ})
of the $\hat{Z}(s,T)$ function (\ref{Zsbag}). Due to this 
there is a possibility of a phase transition in the model.
Whether it does exist and what the order is depends 
on the values of the parameters $\gamma$ and $\delta$ in the pre-exponential
power-like factor, $v^{\gamma}(m-Bv)^{\delta}$, of the mass-volume bag spectrum $\rho(m,v)$
(\ref{rhomv}). 
This resembles the case of the statistical bootstrap model discussed in Sec.~II.
The limiting temperature $T_H=1/b$ appears because of the exponentially increasing factor
$\exp(bm)$, in the mass spectrum (\ref{rhom}), but the thermodynamical behavior
(\ref{a1}-\ref{a3}) at $T=T_H$ depends crucially on the value of the parameter $a$ in
the pre-exponential factor, $m^{-a}$ (\ref{rhom}).

What is the  physical difference between $\alpha >2$ and $1<\alpha\le 2$
in the model? From Eq.~(\ref{sH1}) it follows that the volume
distribution function of quark-gluon bags at $T<T_C$  has the form 
\begin{align}
 W(v) ~\propto~ v^{-\alpha}~\exp\left[-v\left(s_H-s_Q\right)\right]~.
\end{align}
Consequently the average bag volume: 
\begin{align}
 \overline{v}~=~\int_{V_o}^{\infty} dv~v~W(v)~,  
\end{align}
at $T~\rightarrow~(T_{Q}-0)$ approaches:
\begin{align}
& \overline{v}~=~const~,~~~~~~~~{\rm for}~~~~ \alpha~>~2~; \\
& \overline{v}~\rightarrow~\infty~,~~~~~~~~~~~{\rm for}~~~~ 1~<~\alpha
~\le~ 2~. 
\end{align}
Thus in the vicinity of 
the 1$^{st}$ order PT 
the finite volume bags ("hadrons") dominate at
$T< T_C$. 
There is a single infinite volume bag
(QGP) at $T> T_C$ and a mixed phase at $T=T_C$ (see also Ref.~\cite{pt3}).
For the  2$^{nd}$ order and/or "higher order transitions" the dominant bag configurations 
include the large volume bags already in a hadron phase $T<T_C$, and the average
bag volume increases to infinity at $T\rightarrow (T_C-0)$.


The condition for the 2$^{nd}$ order PT can be derived as following.
The integral for the function $F$ in
Eq.~(\ref{GF}) reads
\begin{align}\label{int}
\int_{V_o}^{\infty}dvv^{-\alpha+1}\exp\left[-v\left(s_H-s_Q\right)
 \right]~
 =~\left(s_H-s_Q\right)^{-2+\alpha}~\Gamma\left[2-\alpha,
 (s_H-s_Q)V_o\right]~, 
\end{align}
where $\Gamma(k,x)$ is the incomplete Gamma-function.
Thus using
Eq.~(\ref{sHprime1}) one finds at $T\rightarrow (T_C-0)$:
\begin{align}\label{le2}
 s_Q^{\prime}~-~s_H^{\prime}~&\propto ~ \left(s_H~-~s_Q\right)^{2-\alpha}~,
 ~~~~~~~~~~~~~{\rm for}~~~~~\alpha~<~2~,  \\
 s_Q^{\prime}~-~s_H^{\prime}~&\propto~-~\ln^{-1}\left(s_H~-~s_Q\right)~, 
 ~~~~~~~~{\rm for}~~~~~ \alpha~=~2~,\label{eq2}
\end{align}
and consequently:
\begin{align}
s_H^{\prime\prime}~- s_Q^{\prime\prime}~& \propto ~(s_H~-~s_Q)^{3-2\alpha}~~~~~~~~~
{\rm for}~~~~~~~~~\alpha~<~2~;
\label{*} \\
s_H^{\prime\prime}~- s_Q^{\prime\prime}~& \propto ~-~\frac{\ln^{-3}(s_H-s_Q)}{s_H~-~s_Q}~~~~~~~ 
{\rm for}~~~~~~~~~\alpha~=~2~.
\label{**}
\end{align}
Therefore for $3/2 < \alpha\le 2$ the 2$^{nd}$ order PT
with $ s_H^{\prime\prime}(T_C)=\infty$ takes place whereas
 for $\alpha=3/2$ the 2$^{nd}$ order PT with the finite
value of $ s_H^{\prime\prime}(T_C)$ is observed. 
 The dependence of the specific heat $C\equiv d\varepsilon/dT$ on $T$
is  shown in Fig.~\ref{c2}, left. 

\begin{figure}[h!]
 \hspace{-0.1cm}
\epsfig{file=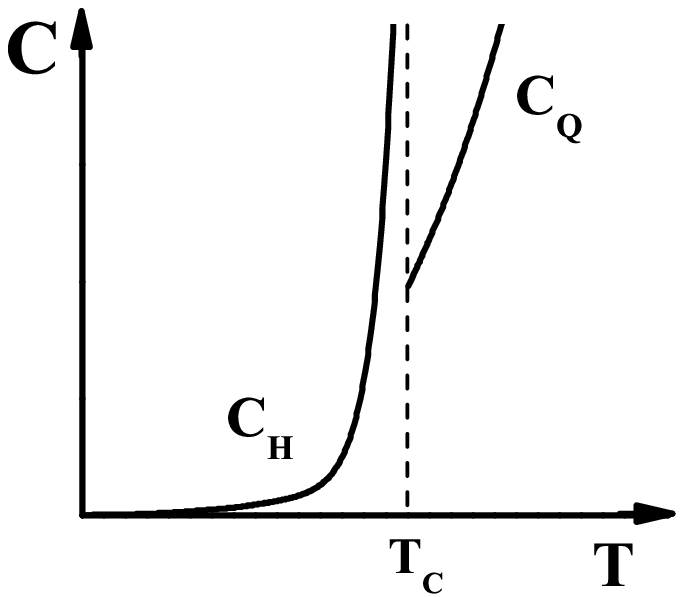,height=8cm,width=7cm} \hspace{1.5cm}
\epsfig{file=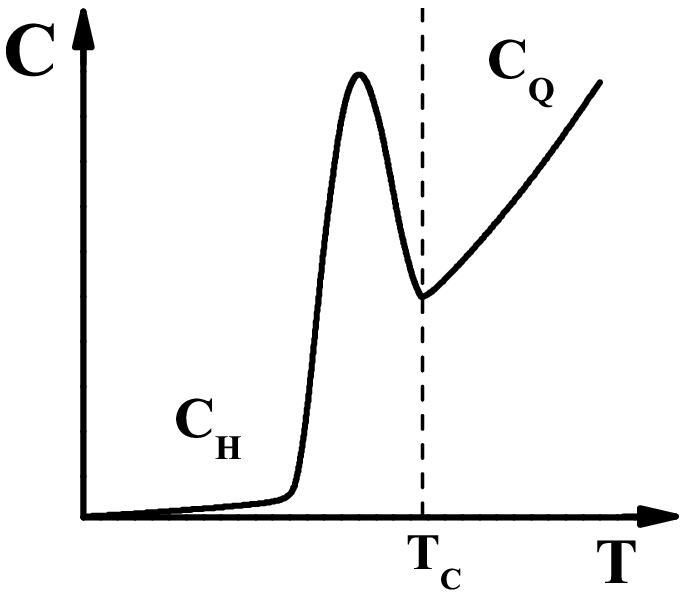,height=8cm,width=7cm}
\caption{ 
The dependence of the
specific heat on temperature for the  2$^{nd}$ order PT (left) and for the $3^{rd}$ order
PT (right). Large values of $C$ in the hadron phase  reflect the fact that large derivative
$d\varepsilon_H/dT$ at $T$ close to $T_C$
is needed for the 2$^{nd}$ and $3^{rd}$ order
PT to reach the value of
$\varepsilon_Q(T_C)$ without energy density discontinuity.  
\label{c2}
}
\end{figure}
An infinite value of a specific heat $C$ at $T\rightarrow (T_C-0)$ 
obtained within the model for
$3/2<\alpha\le 2$  
is typical  for the 2$^{nd}$ order PTs \cite{LL}.


From Eq.~(\ref{*}) it follows that  $s_H^{\prime\prime}(T_C)=s_Q^{\prime\prime}(T_C)$
for $\alpha <3/2$. Using Eqs.~(\ref{le2},~\ref{*}) one finds at $T\rightarrow (T_C-0)$:
\begin{align}\label{***}
s_H^{\prime\prime\prime}~ -~ s_Q^{\prime\prime\prime}~
\propto~ (s_H~-~s_Q)^{4-3\alpha}~.
\end{align}
Thus, for $4/3\le \alpha <3/2$ there is a $3^{rd}$ order transition:
\begin{align}\label{3rd}
s_H(T_C)~ =~ s_Q(T_C)~,~~~~
s_H^{\prime}(T_C)~ =~ s_Q^{\prime}(T_C)~,~~~~
s_H^{\prime\prime}(T_C)~=~ s_Q^{\prime\prime}(T_C)~,~~~~
s_H^{\prime\prime\prime}(T_C)~\ne~ s_Q^{\prime\prime\prime}(T_C)~,
\end{align}
 with $ s_H^{\prime\prime\prime}(T_C)=\infty$
for $4/3 < \alpha< 3/2$ and
 with a finite value of $s_H^{\prime\prime\prime}(T_C)$ for $\alpha=4/3$.
The dependence of the specific heat $C\equiv d\varepsilon/dT$ on temperature
for the $3^{rd}$ order transition
is shown in Fig.~\ref{c2}, right.

By calculating higher order derivatives of $s_H$ and $s_Q$ with respect to $T$
it can be shown that
for $(n+1)/n \le \alpha< n/(n-1)$ ($n=4,5,...$) there is a $n^{th}$ order transition
\begin{align}\label{nth}
s_H(T_C)~ =~ s_Q(T_C)~,~~~~
s_H^{\prime}(T_C)~ =~ s_Q^{\prime}(T_C)~,~~~~...~~~
s_H^{(n-1)}(T_C)~=~ s_Q^{(n-1)}(T_C)~,~~~~
s_H^{(n)}(T_C)~\ne~ s_Q^{(n)}(T_C)~,
\end{align}
 with $ s_H^{(n)}(T_C)=\infty$
for $(n+1)/n < \alpha< n/(n-1)$
and  with a finite value of $s_H^{(n)}(T_C)$ for $\alpha=(n+1)/n$.

The 3$^{rd}$ and higher order PTs  correspond
to a continuous specific heat function $C$ with its maximum at $T$ near $T_C$
(see Fig.~\ref{c2}, right).  
This maximum appears due to 
the fact that large values of derivative
$d\varepsilon_H/dT$ at $T$ close to $T_C$
are needed to reach the value of
$\varepsilon_Q(T_C)$ without discontinuities of energy density and specific heat.  
The so called crossover point is usually defined as a position 
of this maximum. Note that in the present model the maximum of a specific heat is always
inside the hadron phase.

\section{Non-Zero Baryonic Number}
At a non-zero baryonic density the grand canonical partition function
for the system of quark-gluon bags can be presented in the form
 \begin{align}\label{ZVB}
 Z(V,T,\mu_B)~=~
\equiv\sum_{b=-\infty}^{\infty}~\exp\left(\frac{b~\mu_B}{T}\right)
~Z(V,T,B) ~,
\end{align}
where $\mu_B$ is the baryonic chemical potential 
(for simplicity strangeness is neglected in the initial discussion).
The Laplace transform of $Z(V,T,\mu_B)$ (\ref{ZVB}) reads \cite{pt3}
\begin{align}\label{ZsB}
\hat{Z}(s,T,\mu_B)~\equiv~
\int_0^{\infty}dV\exp(-~sV)
Z(V,T,\mu_B)~
=~\frac{1}{s~-~f(s,T,\mu_B)}~,
\end{align}
where
\begin{align}\label{fsB}
f(T,\mu_B,s)~=~f_H(T,\mu_B,s)
~+~\int_{V_o}^{\infty}dv\int_{M_o+Bv}^{\infty}dm~\rho(m,v;\mu_B/T)
~\exp(-sv)\phi(T,m)~,
\end{align}
with
\begin{align}\label{fHTmu}
f_H(T,\mu_B,s)~&=~\sum_{j=1}^n~ g_j~ \exp\left(\frac{b_j \mu_B}{T}\right)~
\exp(-v_js)~\phi(T,m_j)~,\\
\rho(m,v;\mu_B/T)~&\equiv~\sum_{b=-\infty}^{\infty}
 \exp\left(\frac{b\mu_B}{T}\right)~\rho(m,v;b)~.
 \end{align}
Similar to the case of $\mu_B=0$ discussed in the previous sections
one finds that the pressure is defined by the
farthest-right singularity, $s^*$:
\begin{align}\label{psB}
 p(T,\mu_B)
~=~\lim_{V\rightarrow\infty}\frac{T}{V}\ln Z(V,T,\mu_B)~=~T~s^*(T,\mu_B)
 ~=~ T\cdot max\{s_H,~s_Q\}~,
\end{align}
and it can be either given by the pole singularity, $s_H$: 
\begin{align}\label{sHTmu}
s_H(T,\mu_B)~=~f(s_H,T,\mu_B) ~,
\end{align}
or the singularity $s_Q$ of the function $f(s,T,\mu_B)$ (\ref{fsB}) itself:
\begin{align}\label{sQB}
 s_Q(T,\mu_B)~=~ \frac{\pi^2}{90}T^3~ \left[
 \frac{95}{2}~ +~
\frac{10}{\pi^2}~\left(\frac{\mu_B}{T}\right)^2~+~\frac{5}{9\pi^4}~
\left(\frac{\mu_B}{T}\right)^4~\right]
~-~\frac{B}{T} ~
\equiv ~\frac{1}{3}~\overline{\sigma}_Q(\mu_B)~T^3~-~\frac{B}{T}~.
\end{align}
Note that for $\mu_B=0$ Eq.~(\ref{sQB}) is transformed back to $s_Q(T)$ (\ref{sQ}),
as $\overline{\sigma}_Q(\mu_B=0)=\sigma_Q$.
The energy density and baryonic number density are equal to
\begin{align}
 \varepsilon(T,\mu_B)~=
 ~T^2~\frac{\partial s^*(T,\mu_B)}
{\partial T} ~+~T\mu_B~\frac{\partial s^*(T,\mu_B)}
{\partial \mu_B}~,~~~~
 n_B(T,\mu_B)~=
~T~\frac{\partial s^*(T,\mu_B)}
{\partial \mu_B}~ .
\end{align}
   
The second term in Eq.~(\ref{fsB}), the function $f_Q(T,\mu_B,s)$, can be approximated as
\begin{align}\label{fQsB}
 f_Q(T,s,\mu_B)~\simeq~ u(T,\mu_B/T)
 ~  \int_{V_o}^{\infty}dv~v^{-\alpha}~\exp\left[-v\left(s~-s_Q(T,\mu_B)\right)\right]~,
 \end{align}
i.e., it has the same form as  $f_Q(T,s)$ (\ref{fQs}).
At $\mu_B/T=const$ and $T\rightarrow\infty$ one finds $s_Q(T,\mu_B)\propto T^{3}$
and $u(T,\mu_B/T)\propto T^{10+4\delta}$. 
This is the same behavior as in the case of $\mu_B=0$, 
(\ref{fQs}).
%
At a small $T$ and $\mu_B$ one finds $s_H>s_Q$, so that the
farthest-right singularity $s^*$ equals to $s_H$.
This pole-like singularity, 
$s_H(T,\mu_B)$, of the function $\hat{Z}(s,T,\mu_B)$ (\ref{ZsB}) should be compared
with the singularity $s_Q(T,\mu_B)$  of the function $f_Q(T,s,\mu_B)$ (\ref{fsB}) itself.
The dependence of $s_Q$ on the variables $T$ and $\mu_B$
is known in an explicit form (\ref{sQB}).
%
If conditions $(\ref{gamdel})$ are satisfied it can be shown
that with an increasing $T$ along the lines of $\mu_B/T=const$ one 
reaches the point $T_C(\mu_B)$ at which the two singularities
collide, $s_Q=s_H$, and  $s_Q(T,\mu_B)$ becomes the
farthest-right singularity  of the function $\hat{Z}(s,T,\mu_B)$ (\ref{ZsB})
at $T\rightarrow\infty$ and fixed $\mu_B/T$.
Therefore, the line  of  
phase transitions $s_H(T,\mu_B)=s_Q(T,\mu_B)$ appears
in the $T-\mu_B$ plane.
%
%
Below the phase transition
line one observes $s_H (T,\mu_B)>s_Q(T,\mu_B)$ and the system is in a hadron phase.
Above this line the singularity $s_Q(T,\mu_B)$ becomes the
farthest-right singularity  of the function $\hat{Z}(s,T,\mu_B)$ (\ref{ZsB}) and
the system is in a QGP phase.  
The corresponding thermodynamical functions are given by: 
\begin{align}\label{pQB}
p(T,\mu_B)~&=~T~ s_Q(T,\mu_B)~=~\frac{\pi^2}{90}\cdot 
 \frac{95}{2}~T^4~ +~
\frac{1}{9}~\mu_B^2 T^2~+~\frac{1}{162\pi^2}~\mu_B^4~-~B~,\\
\varepsilon(T,y_B)~&=~T^2~\frac{\partial s_Q(T,\mu_B)}{\partial T}~+~T\mu_B~
\frac{\partial s_Q(T,\mu_B)}{\partial \mu_B}~=~
3~ p(T,\mu_B)~+~4B~,\\
n_B(T,\mu_B)~&=~ T~\frac{\partial s_Q(T,\mu_B)}{\partial \mu_B}~=~\frac{2}{9}~\mu_B~T^2~
~+~\frac{2}{81\pi^2} \mu_B^3~.
\end{align}

The analysis similar to that in the previous section leads to 
the conclusion that one has the  1$^{st}$ order PT
for $\alpha > 2$ in Eq.~(\ref{fQsB}), for $3/2 \le \alpha \le 2$ there is the 2$^{nd}$ order PT,
and, in general, for $(n+1)/n \le \alpha< n/(n-1)$ ($n=3,4,...$) there is a $n^{th}$ order transition.
Note that $s_H(T,\mu_B)$ found from by Eq.~(\ref{sHTmu}) is only weakly dependent on $\alpha$. 
This means that for $\alpha >1$ the hadron gas pressure $p_H=Ts_H$ 
and thus the position of the phase transition line,
\begin{align}\label{line}
s_H(T,\mu_B)~=~ s_Q(T,\mu_B)~,
\end{align}
in the $T-\mu_B$ plane is not affected by the contribution from the large volume bags.
The main contribution to $s_H$ (\ref{sHTmu}) comes from small mass (volume)
bags, i.e. from known hadrons included in $f_H$ (\ref{fHTmu}). This is valid for all $\alpha >1$,
so that the line (\ref{line}) calculated within the model is similar 
for transitions of different orders.  
On the other hand, the behavior of the derivatives of $s_H$ (\ref{sHTmu}) with respect to $T$ and/or
$\mu_B$ near the critical line (\ref{line}), and thus the order of the phase transition,  
may  crucially depend on the contributions from the quark-gluon bags with $v\rightarrow\infty$. 
For $\alpha >2$ one observes the 1$^{st}$ order PT, but for $1<\alpha\le 2$ the  2$^{nd}$ and higher
order PTs are found. In this latter case the energy density, baryonic number density and entropy density
have significant contribution from the large volume bags in the hadron phase
near the PT line (\ref{line}).  

The actual structure of the "critical" line on the $T-\mu_B$ plane
is defined by a dependence of the parameter $\alpha$ on the $\mu_B/T$ ratio.
This dependence can not be reliably evaluated within the model and
thus an external information is needed in order to locate
the predicted "critical" line in the phase diagram.
The lattice QCD calculations indicate \cite{misha} that at zero $\mu_B$
there is rapid but smooth cross-over.
Thus this suggests a choice $1 < \alpha < 3/2$ at $\mu_B = 0$,
i.e. the transition is of the 3$^{rd}$ or a higher order. 
Numerous models predict the strong 1$^{st}$ order PT at a high
$\mu_B/T$ ratio \cite{misha}, thus $\alpha > 2$ should be selected in
this domain.  
As a simple example in which the above conditions are
satisfied one may  consider a linear dependence,
$\alpha = \alpha_0+\alpha_1\mu_B/T$,
where $\alpha_0=1+\epsilon$ ($0<\epsilon<<1$) and $\alpha_1\approx 0.5$.
Then the line of the 1$^{st}$ order PT at a high $\mu_B/T$ 
ends at the point $\mu_B/T \approx 2$, where
the line
of  the 2$^{nd}$ order PT starts.
Further on at $\mu_B/T\approx 1$ the lines of the 3$^{rd}$ and higher order transitions follow on the
"critical" line.
This hypothetical "critical" line of the deconfinement phase transition in
the $T-\mu_B$ plane is shown in 
Fig.~\ref{tmu}.

\begin{figure}[h!]
 \hspace{-0.1cm}
\epsfig{file=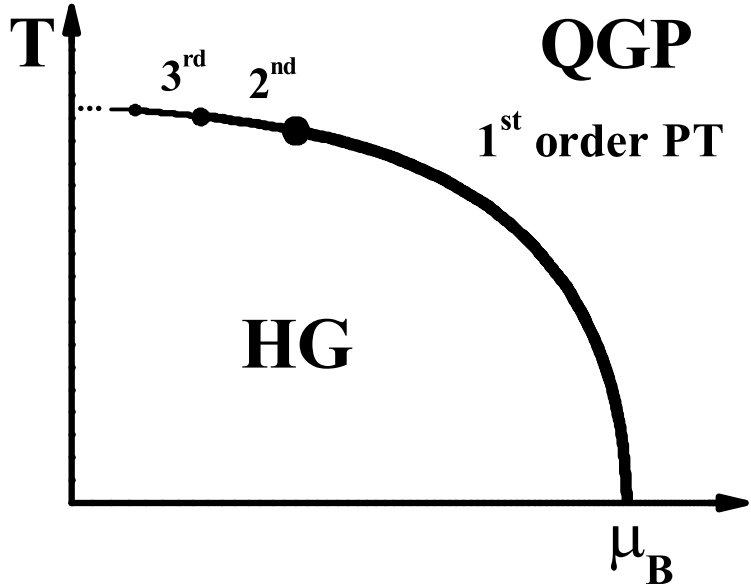,height=8cm,width=10cm}
\caption{
The hypothetical phase diagram of strongly interacting matter in the $T-\mu_B$
plane  within the quark-gluon bag model.  The influence of strangeness
is neglected. The line of the 1$^{st}$ order phase transition at a
high $\mu_B/T$ ratio is followed by the line of the
2$^{nd}$ order PT at an intermediate $\mu_B/T$ values and by the lines
of higher order PTs at a low $\mu_B/T$. 
\label{tmu}
}
\end{figure}

In the case of
the non-zero strange chemical potential $\mu_S$ the pole singularity,
$s_H$, and the singularity $s_Q$ become dependent on $\mu_S$.
The system created in nucleus-nucleus collisions has zero net strangeness
and consequently,
\begin{align}\label{S0}
n_S(T,\mu_B,\mu_S)~=
T~\frac{\partial s^*(T,\mu_B,\mu_S)} {\partial
\mu_S}~=~0~.
\end{align}
At a small $T$ and $\mu_B$, when $s_H>s_Q$, Eq.~(\ref{S0}) with $s^*=s_H$ defines
the strange chemical potential $\mu_S=\mu_S^H(T,\mu_B)$ which guarantees
a zero value of the net strangeness density in a hadron phase. 
When the singularity $s_Q$ becomes the 
farthest-right singularity the requirement of zero net strangeness (\ref{S0})
with $s^*=s_Q$ leads to $\mu_S=\mu_S^Q=\mu_B/3$. The functions $\mu_S^H(T,\mu_B)$
and $\mu_S^Q(T,\mu_B)=\mu_B/3$ are  different. Consequently, the line
(\ref{line}) of the  1$^{st}$ order PT  in the $T-\mu_B$ plane is transformed 
into a "strip"~\footnote{Within 
a phenomenological construction of the 1$^{st}$
order PT this was discussed in Ref.~\cite{heinz}.}. 
The phase diagram in which influence of the strangeness is taken into
account is shown schematically in Fig.~\ref{tmu2}.

\begin{figure}[h!]
 \hspace{-0.1cm}
\epsfig{file=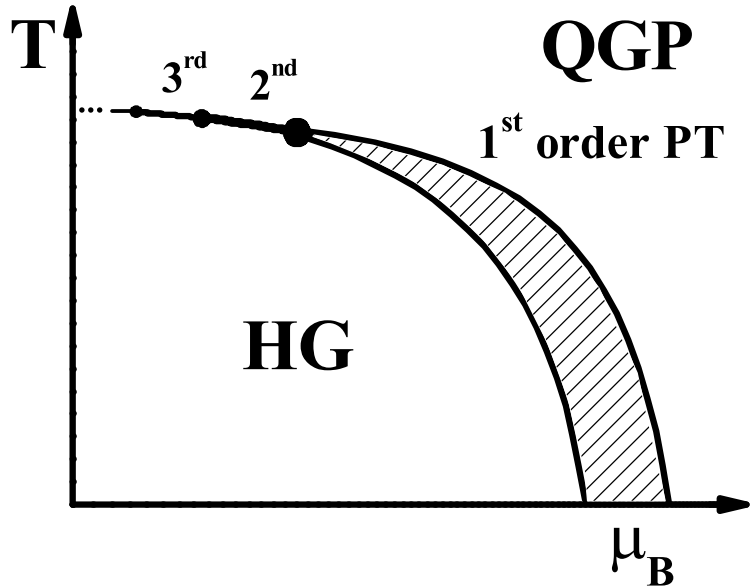,height=8cm,width=10cm}
\caption{ The hypothetical phase diagram of strongly interacting matter in the $T-\mu_B$
plane with strangeness  taken into
account.  The line of the 1$^{st}$ order phase transition is transformed 
into a "strip". For further details see caption of Fig.~\ref{tmu}.
\label{tmu2}
}
\end{figure}
The lower and upper limits of the "strip"  are defined, respectively, by the conditions:
\begin{align}\label{low-up}
s_H(T,\mu_B,\mu_S^H)~=~s_Q(T,\mu_B,\mu_S^H)~,~~~~s_Q(T,\mu_B,\mu_S^Q)~=~s_H(T,\mu_B,\mu_S^Q)~.
\end{align}
The $T$ and $\mu_B$ values inside the "strip"
correspond to the mixed HG-QGP phase for which the 
following conditions have to be satisfied:
\begin{align}\label{mix1}
&s_H(T,\mu_B,\mu_S)~=~s_Q(T,\mu_B,\mu_S)~,\\
&n_S^{mix}~\equiv~\delta \cdot~n_S^Q~+~
(1 ~-~\delta)\cdot~ n_S^H~=
~\delta\cdot~T~\frac{\partial s_Q} {\partial
\mu_S}~+~(1-\delta)\cdot~T~\frac{\partial s_H} {\partial
\mu_S}~=~0~, \label{mix2}
\end{align}
where $\delta$ is a fraction  of the mixed phase volume occupied
by the QGP.
Eq.~(\ref{mix1}) reflect the Gibbs equilibrium between
the two phases: the thermal equilibrium (equal temperatures), the mechanical equilibrium
(equal pressures), the chemical equilibrium (equal chemical potentials). The net strangeness
does not vanish in each phase separately, but the total net strangeness 
of the mixed phase (\ref{mix2}) is equal to zero. This means the strangeness--anti-strangeness
separation inside the mixed phase \cite{GKS}.  
The line (\ref{line}) for the 2$^{nd}$ and higher order PTs (i.e. for $1<\alpha\le2$)
remains unchanged as $\mu_S^H$ becomes equal to $\mu_S^Q$
along this line.

\section{Summary and Outlook}

The phase diagram of strongly interacting matter was studied 
within the exactly solvable 
statistical model of the quark-gluon bags. The model predicts two phases
of matter: the hadron gas at a low temperature $T$ and baryonic chemical potential
$\mu_B$, and the quark-gluon gas at a high $T$ and/or $\mu_B$. The order of the phase transition
is expected to alter with a change of the $\mu_B/T$ ratio.
The line of the 1$^{st}$ order transition at a high $\mu_B/T$
 is followed by the line
of  the 2$^{nd}$ order phase transition at an intermediate $ \mu_B/T$, and then by the lines
of the "higher order transitions" at a low $\mu_B/T$. 
The condition of the strangeness conservation transforms the 1$^{st}$ order transition
line into the "strip" in which the strange chemical potential varies between
the QGP and HG values.

In the high and low temperature domains
the approach presented here reduces to two well known and
successful  models:
the hadron gas model and the bag model of QGP.
Thus, one may hope the obtained results concerning properties
of the phase transition region may reflect the basic features of
nature.

Clearly, a further development of the model is possible and
required. It is necessary to perform quantitative
calculations using known hadron states.
These calculations should allow to establish a relation between
the model parameters and the $T$ and $\mu_B$ values  and
hence locate the critical line in the $T-\mu_B$ plane.
Further on one should investigate various possibilities
to study experimentally a rich structure of the transition
line predicted by the model.

\vspace{1cm}
{\bf Acknowledgments} \\
We are grateful to V.V.~Begun, C.~Greiner, E.V.~Shuryak, M.~Stephanov, and I.~Zakout for useful comments
and Marysia Gazdzicka for corrections. 
\noindent
The work was supported by US Civilian Research
and Development Foundation (CRDF) Cooperative Grants Program,
Project Agreement UKP1-2613-KV-04 and
by the Virtual Institute VI-146
of Helmholtz Gemeinschaft, Germany.

\end{document}